# Decoherence-protected quantum gates for a hybrid solid-state spin register

T. van der Sar[1], Z. H. Wang[2], M. S. Blok[1], H. Bernien[1], T. H. Taminiau[1], D.M. Toyli[3], D. A. Lidar[4], D. D. Awschalom[3], R. Hanson[1], V. V. Dobrovitski[2*]

1. Kavli Institute of Nanoscience, Delft University of Technology, P.O. Box 5046, 2600 GA Delft, The Netherlands

2. Ames Laboratory and Iowa State University, Ames, Iowa 50011, USA

3. Center for Spintronics and Quantum Computation, University of California, Santa Barbara, California 93106, USA

4. Departments of Electrical Engineering, Chemistry, and Physics, and Center for Quantum Information Science and Technology, University of Southern California, Los Angeles, California 90089, USA

**Protecting the dynamics of coupled quantum systems from decoherence by the environment is a key challenge for solid-state quantum information processing[1-3]. An idle qubit can be efficiently insulated from the outside world via dynamical decoupling[4], as has recently been demonstrated for individual solid-state qubits[5-10]. However, protection of qubit coherence during a multi-qubit gate poses a non-trivial problem[4,11,12]: in general the decoupling disrupts the inter-qubit dynamics, and hence conflicts with gate operation. This problem is particularly salient for hybrid systems[13-23], wherein different types of qubits evolve and decohere at vastly different**

**rates. Here we present the integration of dynamical decoupling into quantum gates for a paradigmatic hybrid system, the electron-nuclear spin register. Our design harnesses the internal resonance in the coupled-spin system to resolve the conflict between gate operation and decoupling. We experimentally demonstrate these gates on a two-qubit register in diamond operating at room temperature. Quantum tomography reveals that the qubits involved in the gate operation are protected as accurately as idle qubits. We further illustrate the power of our design by executing Grover's quantum search algorithm[1], achieving fidelities above 90% even though the execution time exceeds the electron spin dephasing time by two orders of magnitude. Our results directly enable decoherence-protected interface gates between different types of promising solid-state qubits. Ultimately, quantum gates with integrated decoupling may enable reaching the accuracy threshold for fault-tolerant quantum information processing with solid-state devices[1,12].**

Decoherence is a major hurdle towards realizing scalable quantum technologies in the solid state. The inter-qubit dynamics that implement the quantum logic are unavoidably affected by uncontrolled couplings to the solid-state environment, preventing high-fidelity gate performance (Fig 1a). Dynamical decoupling[4], a technique that employs fast qubit flips to average out the interactions with the environment, is a powerful and practical tool for mitigating decoherence[5-12,24,25]. This approach is particularly promising for the emerging class of hybrid quantum architectures[13-23] in which different types of qubits, such as electron and nuclear spins, superconducting resonators, and nanomechanical oscillators, perform different functions. Dynamical decoupling allows for each qubit type to be decoupled at its own appropriate rate, ensuring uniform coherence protection.

However, combining dynamical decoupling with quantum gate operations is generally problematic, since decoupling does not distinguish the desired inter-qubit interaction from the coupling to the decohering environment, and in general cancels both (Fig. 1b). For hybrid systems, where large difference in coherence and control timescales among the different qubit types make the encoding-based schemes[11] or synchronized application of decoupling pulses[4,12] fail, a solution has thus far remained elusive.

Here we present a design that enables the integration of decoupling into gate operation for hybrid quantum architectures. We demonstrate such decoherence-protected gates on a prototype hybrid quantum system: a two-qubit register consisting of an electron and a nuclear spin (Fig. 1c). The key idea is to precisely adapt the time intervals between the electron decoupling pulses to the nuclear spin dynamics. When combined with continuous nuclear spin driving, this synchronization yields selective rotations of the nuclear spin while the electron spin is dynamically protected, as explained below. This design preserves all of the advantages of dynamical decoupling without requiring additional qubits or controllable inter-qubit couplings. It can be readily implemented to yield decoherence-protected quantum gates in a range of hybrid systems, such as various electron-nucleus spin registers[13-17,20], and interface gates between the qubits and a spin-chain quantum databus[22,23].

We experimentally demonstrate the scheme on a single nitrogen-vacancy (NV) center in diamond[13,16,26-28], where the two qubits are represented by the electron spin and the host $^{14}$N nuclear spin (see Fig.1d and supplementary information). An entangling gate between these qubits can be implemented using the hyperfine interaction, described by the Hamiltonian $\hat{H}_{hf} = A\hat{I}_Z\hat{S}_Z$, where $A = 2\pi \cdot 2.16$ MHz for NV centers[27,28] and $\hat{I}_Z(\hat{S}_Z)$ is the nuclear (electronic) spin operator. For an unprotected gate, of duration $T_G = \pi/A$, the fidelity is limited by the electron spin dephasing, which is dominated either by the bath of $^{13}$C nuclear spins[13] (as in the experiments here) or by the electron spins of substitutional N atoms[26]. Decoupling applied to the electron spin suppresses decoherence, but also disrupts the electron-nuclear coupling. At the same time, synchronized application of the decoupling pulses to both the electronic and the nuclear spin qubit is problematic, because a nuclear spin rotation takes longer (>10 µs for realistic Rabi driving[17,27]) than the electron spin's dephasing time $T_2^*$ (0.5-5 µs for NV centers[13,26]).

Within our gate design, Rabi driving is applied at the frequency of the nuclear spin transition corresponding to the electron spin state $|1\rangle$ (Fig. 1d). In the rotating frame, the two spins are described by the Hamiltonian $\hat{H} = A\hat{I}_Z\hat{S}_Z + \omega_1\hat{I}_X$, where $\omega_1$ is the nuclear Rabi frequency. The nuclear spin dynamic is then conditioned on the electron spin state (Fig. 1e and supplementary information): the nuclear spin undergoes driven rotation around the x-axis if the electron spin is in $|1\rangle$, and precesses around z if the electron is in $|0\rangle$. Although Fig. 1d shows the case of $\omega_1 \ll A$, relevant to our experiments, the scheme works for arbitrary $\omega_1$ and $A$ (see supplementary information).

In parallel with driving the nuclear spin, we decouple the electron spin from the environment with short microwave pulses that constantly switch the electron spin between $|0\rangle$ and $|1\rangle$ states[5,6,8]. The full decoupling sequence consists of repeating the basic unit $\tau$-X-$2\tau$-Y-$\tau$ (see Fig. 2a), where X and Y are decoupling pulses that flip the electron spin around the *x* and *y* axis respectively, and $2\tau$ is the inter-pulse delay. To implement an entangling gate, the nuclear spin dynamics over the full gate unit must depend only on the initial state of the electron spin and cannot be disrupted by the electron's fast switching. This is achieved by setting $\tau = (2n+1)\pi/A$ with integer *n*, so that the nuclear spin rotates by 180º around the *z*-axis during the interval $\tau$, if the electron is in the state $|0\rangle$ during this time. On the other hand, if the electron is in the state $|0\rangle$ during the time interval $2\tau$, the nuclear spin undergoes a 360º rotation that has zero net effect. The overall dynamics during one full gate unit are visualized in Fig. 2b: depending on the initial state of the electron, the nuclear spin rotates around the *x*-axis by the angle $\theta = 4\omega_1\tau$ in the clockwise or counter-clockwise direction (conditional rotation). An unconditional rotation of the nuclear spin, independent of the electron spin state, is constructed from the same gate unit by choosing $\tau = 2n\pi/A$ (Supplementary Fig. 3). From the conditional and unconditional rotations we can construct a complete set of gates for the two-qubit register.

We implement the controlled-rotation gate using a conditional nuclear spin rotation followed by an unconditional rotation over the same angle. The experimental data in Fig. 2c confirm the selectivity of this gate: as the number of gate units increases, the nuclear spin rotates around *x* if the electron spin is initially in $|1\rangle$, and does nothing if the electron spin is initially in $|0\rangle$. When the total nuclear rotation angle equals 180º, the gate corresponds to the controlled-NOT (CNOT) gate (up to an electron phase of $\pi/2$). To fully characterize the CNOT gate, we performed quantum process tomography of its action (Fig. 2d). We find an overall process fidelity $F_p = \text{Tr}\chi_{\text{ideal}}\chi = 83(1)\%$ where $\chi$ is the measured process matrix, and $\chi_{ideal}$ is the matrix of the ideal CNOT gate. This number is a lower bound on the true process fidelity, as it includes errors from imperfect initialization (estimated to be about 5% for this experiment) and errors in the pulses used for state preparation and readout (estimated to be a few %), see supplementary information.

A crucial step of this work is testing our design in the presence of stronger decoherence, and confirming that it ensures efficient protection during the gate operation. In order to change the level of decoherence in a controllable manner, we inject low-frequency noise into the system[24], thus shortening the electron spin-echo decay time $T_2$ from 251(7) μs to 50(2) μs (Fig. 3a). Furthermore, we reduce the nuclear driving power such that the gate time becomes 120 μs, more than twice $T_2$.

We first verify that the additional decoherence can be efficiently suppressed at the single-qubit level by dynamical decoupling. We observe that the electron spin coherence time is extended from 50(2) μs to 234(8) μs as the number of the decoupling pulses is increased from one (spin echo) to sixteen (Fig. 3a), in agreement with previous studies[5,6,8]. Then, we study the fidelity of the CNOT gate as a function of the number of decoupling pulses applied during its 120 μs gate time. The resonance requirements for $\tau$ are maintained by decreasing $n$ for increasing number of pulses (Fig. 3b). We apply the gate to the state $(|0\rangle + i|1\rangle) \otimes |\uparrow\rangle$; this ideally yields the entangled state $|\Psi^+\rangle = |0\uparrow\rangle + |1\downarrow\rangle$. Quantum state tomography reveals that the coherence of the output state, visible as off-diagonal elements in the density matrix $\rho$, grows rapidly with the number of pulses (Fig. 3c). The state fidelity[1] $F = \sqrt{\langle\Psi^+|\rho|\Psi^+\rangle}$ reaches 96(1)% for 16 decoupling pulses (Fig. 3d). Here, the gate performs similarly as without the introduced decoherence (dashed lines in Fig. 3d), showing that the gate efficiency remains high even in the regime where the gate duration exceeds $T_2$. Moreover, comparison with single-qubit state fidelities under decoupling (taken from Fig. 3a) demonstrates that the electron spin coherence in the course of the gate operation is preserved as efficiently as for an idle electron qubit (Fig. 3d). Numerical simulations provide evidence that the fidelity of the CNOT gate remains high even in the presence of much stronger decoherence caused by N impurities[5,26] (see supplementary information), which is highly relevant for scalable quantum architectures in which chains of N atoms serve as databuses between NV centers[22,23].

Finally, we illustrate the power of the decoherence-protected gates by implementing a two-qubit algorithm on a hybrid solid-state spin register. We execute Grover's quantum algorithm[1] for searching a given entry in an unstructured list of $L$ elements. A classical search presents each entry in turn to an "oracle", which outputs 1 if this is the target entry and 0 otherwise. This requires ~ $L/2$ oracle calls on average. The quantum algorithm encodes each entry as a state of an $m$-qubit system ($2^m \geq L$), and presents a superposition of all entries to an oracle. The result is processed to increase the weight of the target state in the superposition. After $O(\sqrt{L})$ iterations, this weight becomes close to 1. For two qubits ($L=4$), a single oracle call already provides an exact answer.

Fig. 4a shows all the pulses that implement the algorithm. The total execution time is 322 µs. The circuit diagram of the full computation is given in Fig. 4b for the target state $|1\downarrow\rangle$, with the conditional phase (CPhase) gate presented separately in Fig.

4c. Figure 4d shows quantum tomographic snapshots of the corresponding two-qubit states at different stages of the algorithm[29]. The fidelity of the resulting state is 95(1) %; for the other target states the fidelities are 92(1) % ($|0\uparrow\rangle$), 91(2) % ($|0\downarrow\rangle$), and 91(1) % ($|1\uparrow\rangle$) (see supplementary information).

Our demonstration of decoherence-protected quantum gates based on resonantly applied decoupling pulses clears the way towards high-fidelity transfer, processing and retrieval of quantum information in small quantum registers, critical tasks for future quantum repeaters and quantum computers[1-3,14-17,30]. Moreover, our gate design is compatible with quantum error correction, and therefore marks an important step towards scalable fault-tolerant quantum computation in a hybrid qubit architecture.

## METHODS

**System initialization and readout.** All experiments are performed at a magnetic field of ~510 G aligned parallel to the NV center symmetry axis. The two-qubit system is initialized into the $(m_S, m_I) = (0,+1)$ state by 4 μs green (532 nm) laser excitation (details in supplementary information). Electron spin polarization into the $m_S = 0$ state is a result of a spin-dependent relaxation mechanism between the electronic excited state and electronic ground state. At 510 G a level anti-crossing in the electronic excited state enables electron-nuclear spin flip-flops, which, in combination with the mechanism responsible for electron spin polarization, leads to nuclear spin polarization into the $m_I = +1$ state. The state of the system is read out by counting the spin-dependent number of photons emitted into the phonon-sideband upon green laser excitation during a detection time window of 1-2 μs (details in supplementary information)[27,28].

**Influence of longitudinal relaxation of the electron spin.** All state fidelity numbers include errors due to spin relaxation within and out of the two-qubit subspace. From independently measured relaxation rates, we estimate that relaxation out of (within) the two-qubit subspace is 0.5% (0.4%) for Fig. 2d, 1% (0.8%) for Fig. 3, and 2.9% (2.1%) for Fig. 4 (details in supplementary information).

**Supplementary Information** is linked to the online version of the paper at www.nature.com/nature.

**Acknowledgments** We thank L. DiCarlo, M. D. Lukin, and L. M. K. Vandersypen for useful discussions and comments. T.v.d.S., H.B. and R.H. acknowledge support from the Dutch Organization for Fundamental Research on Matter (FOM) and the Netherlands Organization for Scientific Research (NWO). D.D.A. and R.H. acknowledge support from DARPA QuEST, and D.D.A. from AFOSR and ARO MURI. D.A.L. was sponsored by the National Science Foundation under grant numbers CHM-924318 and CHM-1037992. Work at Ames Laboratory was supported by the Department of Energy – Basic Energy Sciences under Contract No. DE-AC02-07CH11358.

**Author Contributions** Z.H.W., D.A.L., and V.V.D. performed the gate design and the theoretical analysis. H.B. and D.M.T. performed device fabrication. T.v.d.S., M.S.B., T.H.T., D.D.A., and R.H. designed and performed the experiments. T. v.d.S., R.H. and V.V.D. wrote the manuscript. All authors discussed the results and commented on the manuscript.

**Author information** Reprints and permissions information is available at www.nature.com/reprints. The authors declare no competing financial interests. Correspondence and requests for materials should be addressed to V.V.D. (slava@ameslab.gov).

**Figure captions.**

**Figure 1. Quantum gate operation in the presence of decoherence**

**a-c.** Challenge of high-fidelity quantum gates for qubits (orange: electron spin and purple: nuclear spin) coupled to a decohering environment. **a.** Without decoherence protection, the fidelity of two-qubit gates is limited by interactions with the environment. **b.** Dynamical decoupling efficiently preserves the qubit coherence (protected storage) by turning off the interaction between the qubit and its environment. However, this generally also decouples the qubit from other qubits and prevents two-qubit gate operations. If the decoupling and the gate are separated in time, the unprotected gate is still susceptible to decoherence-induced errors. **c.** The goal is to perform dynamical decoupling during the gate operation, thus ensuring decoherence-protected gates. The gate operation should therefore be compatible with decoupling. The dephasing rate of the nuclear spin is negligible in our experiments. However, nuclear spin protection can easily be incorporated using another layer of decoupling.

**d.** The two-qubit system used in this work: a nitrogen-vacancy (NV) colour centre in diamond carries an electron spin $S = 1$ (orange) coupled to a $^{14}$N nuclear spin $I = 1$ (purple). The states of the electronic qubit $|0\rangle$ and $|1\rangle$ are split by 1.4 GHz in an external field $B_0$=510 G. The states $|0\uparrow\rangle$ and $|0\downarrow\rangle$ are split by 5.1 MHz due to nuclear quadrupolar and Zeeman interactions. The hyperfine coupling yields an additional splitting, so that the levels $|1\uparrow\rangle$ and $|1\downarrow\rangle$ are separated by 2.9 MHz. The Rabi driving is applied in resonance with this transition.

**e.** Dynamics of the electron-nuclear spin system in the limit $\omega_1 \ll A$, visualized in a coordinate frame that rotates with the frequencies 1.4 GHz in the electron spin subspace and 2.9 MHz in the nuclear spin subspace. In this frame, the states $|1\uparrow\rangle$ and $|1\downarrow\rangle$ have the same energy. The Rabi driving field, directed along the $x$ axis, coherently rotates the nuclear spin if the electronic qubit is in $|1\rangle$. On the other hand, the Rabi driving is negligible for the states $|0\uparrow\rangle$ and $|0\downarrow\rangle$ which differ in energy by $A = 2\pi \cdot 2.16$ MHz. The phase accumulation between the states $|0\uparrow\rangle$ and $|0\downarrow\rangle$ corresponds to a coherent rotation of the nuclear spin around the $z$ axis with frequency $A$.

**Figure 2. Concept and demonstration of decoherence-protected quantum gates for an electron-nuclear spin register.**

**a.** Basic unit of the decoherence-protected gate, consisting of dynamical decoupling of the electron spin and continuous nuclear spin driving. X (Y) denotes electron π-pulses around the $x$ ($y$) axis, and τ is the delay time. **b.** Bloch spheres visualize the nuclear spin dynamics, conditioned on the initial electronic spin state, during a single gate unit. **c.** Decoherence-protected controlled-rotation of the nuclear spin. Plotted is the measured probability of the nuclear spin to be in the $|\uparrow\rangle$ state as a function of the number of applied gate units, for the two different electron spin input states. Lines are fits taking into account longitudinal relaxation of the electron spin which was measured independently. For details of the fitting and dynamics of other input states, see Supplementary Information. **d.** Measured process matrix $\chi$ of the decoherence-protected CNOT gate. The transparent bars indicate the values of the matrix elements for the ideal gate.

**Figure 3. Performance of the CNOT gate in the presence of strong decoherence.**

**a.** Dynamical decoupling of the electron spin in the presence of artificially generated low-frequency noise, which is injected into the sample via the same microwave stripline used for the qubit control pulses. The coherence time of the electron spin is effectively prolonged by increasing the number of decoupling pulses $N$ in the applied microwave pulse sequence $\pi/2$-$(\tau$-$\pi$-$\tau)^N$-$\pi/2$, where $\tau$ is the free evolution time. The dots are experimental data, the lines are fits. Fit details are given in the supplementary information. The inset shows the power spectrum of the injected noise (log scale).
**b.** Pulse sequences used to test the performance of the CNOT gate with increasing number of decoupling pulses $N$ and correspondingly decreasing interpulse delay $\tau$ (total gate time is fixed at 120 μs). **c.** Density matrices (real part) of the states created with the pulse sequences shown in panel (b). A -$\pi/2$ rotation over the $x$-axis on the electron prepares the state $(|0\rangle + i|1\rangle) \otimes |\uparrow\rangle$, which is transformed into the Bell state $|0\uparrow\rangle + |1\downarrow\rangle$ by a 120 μs CNOT gate. **d.** Fidelity, overlap, and concurrence of the states shown in (**c**) as a function of the number of decoupling pulses in the CNOT gate. The dots are measured data connected by solid lines. The dashed horizontal lines correspond to the performance of the same gate with 16 decoupling pulses in the absence of the artificially generated noise.. For comparison, the state overlap for decoupling of an idle electron spin, not involved in gate operation, is also shown (black squares and line). All error bars are ± 1 s.d.

**Figure 4. Grover's search algorithm executed with decoherence-protected gates.**

**a.** Pulse sequence implementing Grover's algorithm for the target state $|1\downarrow\rangle$. All nuclear spin rotations are implemented with decoherence-protected gates. Each nuclear $\pi/2$ rotation is performed using two gate units. The electron and nuclear spin rotation axes are adjusted by changing the phase of the driving field. The total computation time is 322 µs, exceeding the electron spin echo decay time $T_2$=251 µs by about 30%. **b.** Quantum circuit diagram of Grover's algorithm for the target state $|1\downarrow\rangle$, corresponding to the pulse sequence shown in panel (a).

**c.** Quantum circuit diagram of the controlled-phase (CPhase) gate, implemented by embedding a CNOT gate in between two nuclear-rotation $\pi/2$ gates that rotate the nuclear spin basis. The pulses on the electron correct the single-qubit phase shift.

**d.** Density matrices of the two-qubit system at different stages of the Grover's algorithm. Stage (i) corresponds to the initial superposition of all 2-qubit states, stage (ii) is the oracle's output, and stage (iii) is the final state of the register, see also panels (a) and (b). The fidelity of the final state is 95(1) %, mainly limited by longitudinal relaxation of the electron spin.

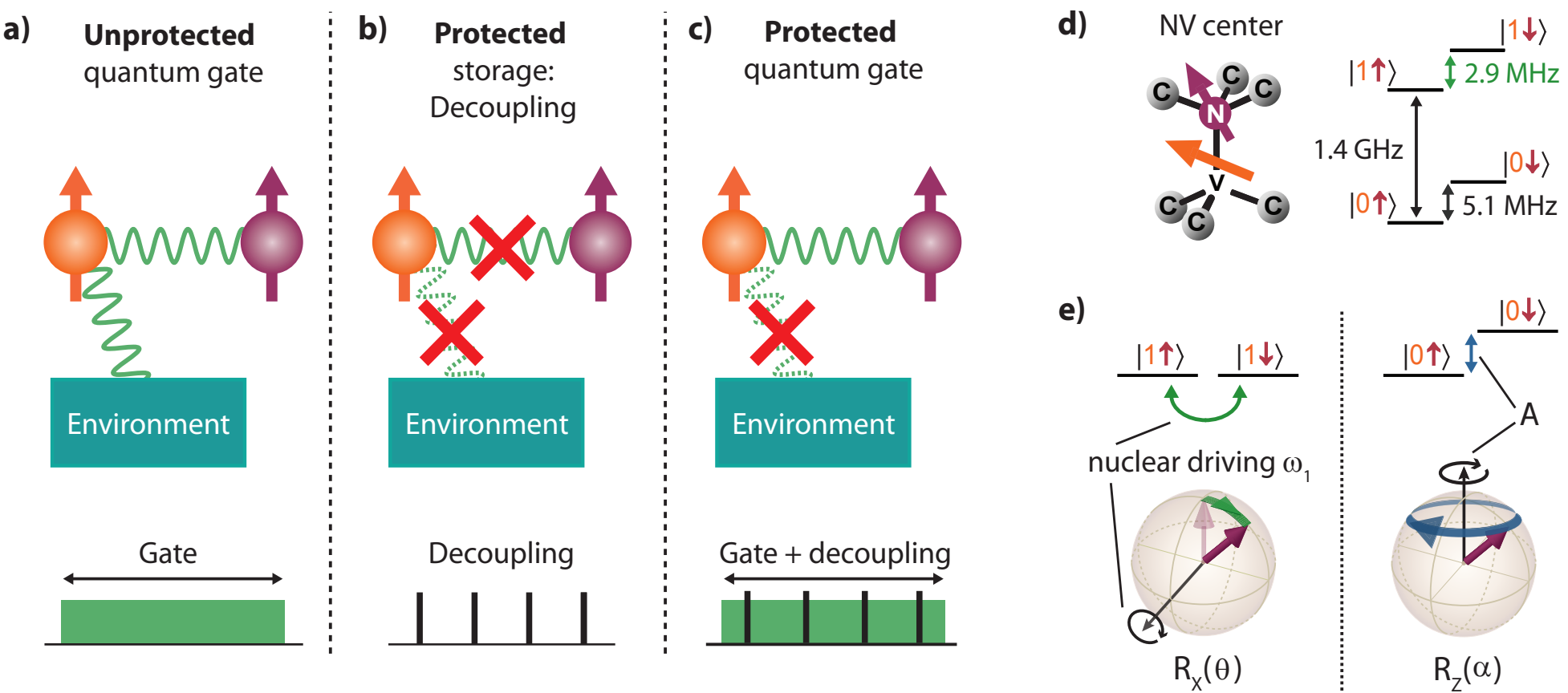

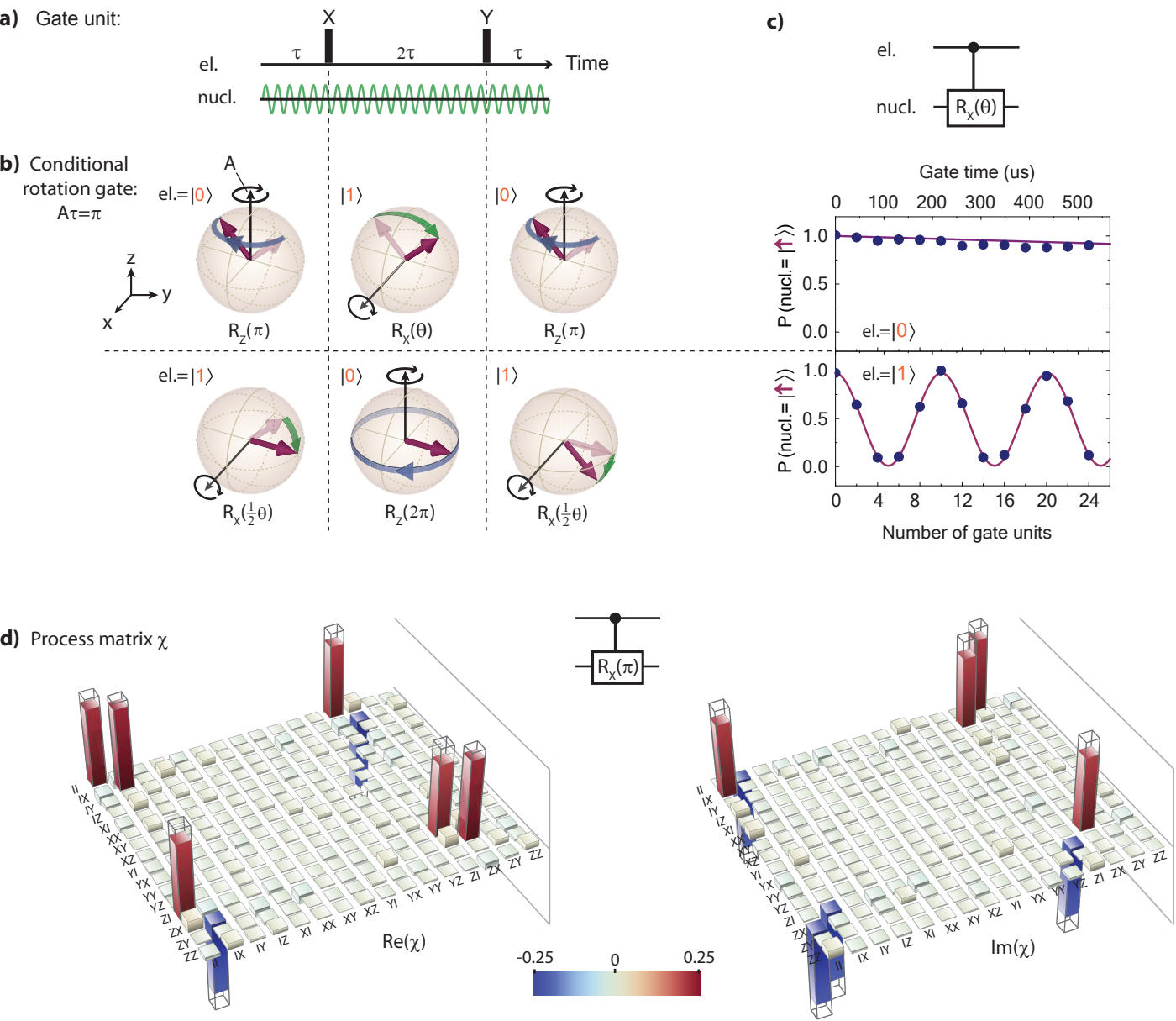

Dobrovitski, V.V., Figure 2

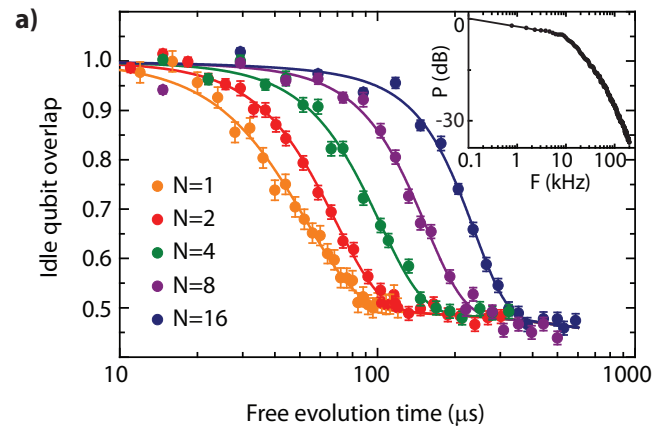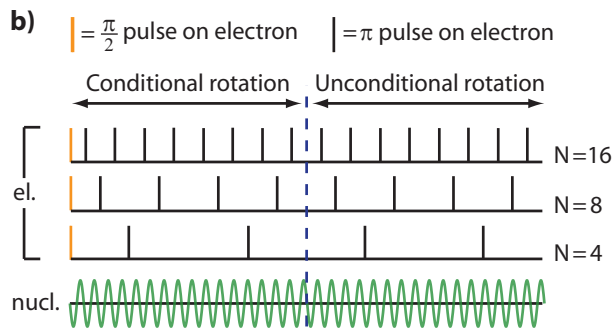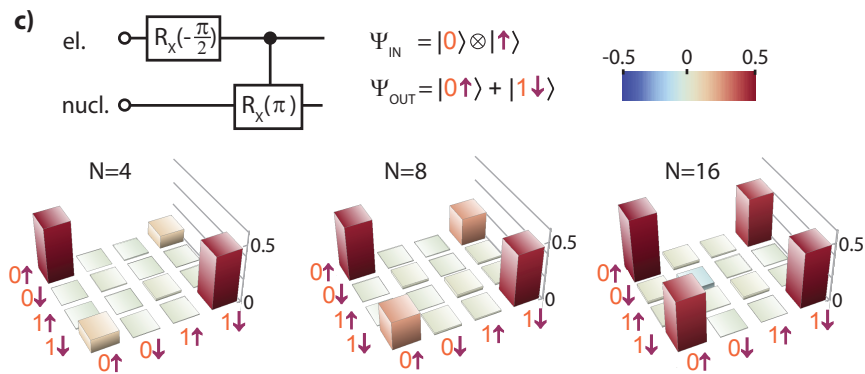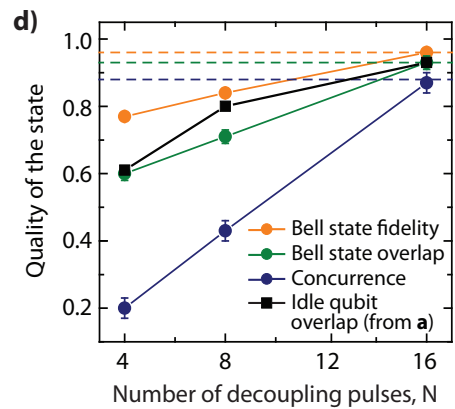

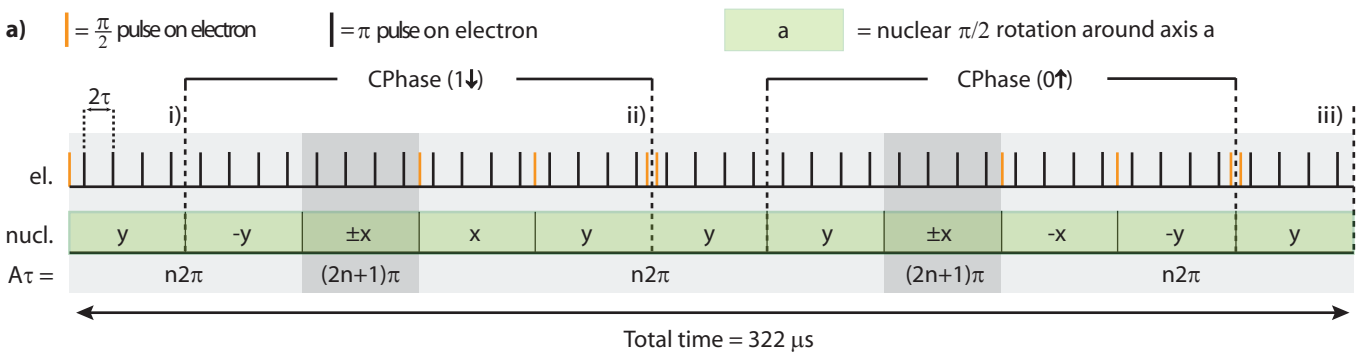
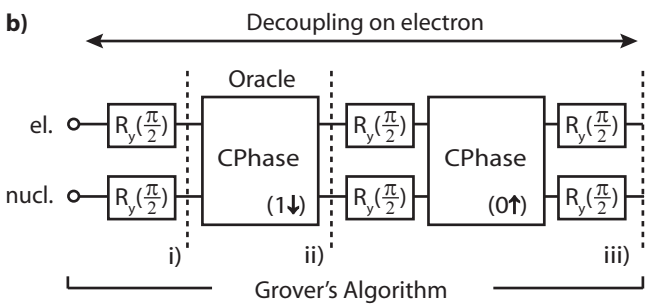
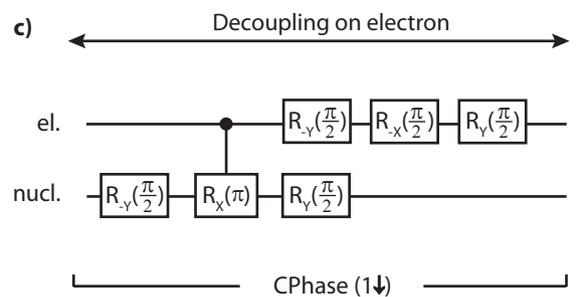
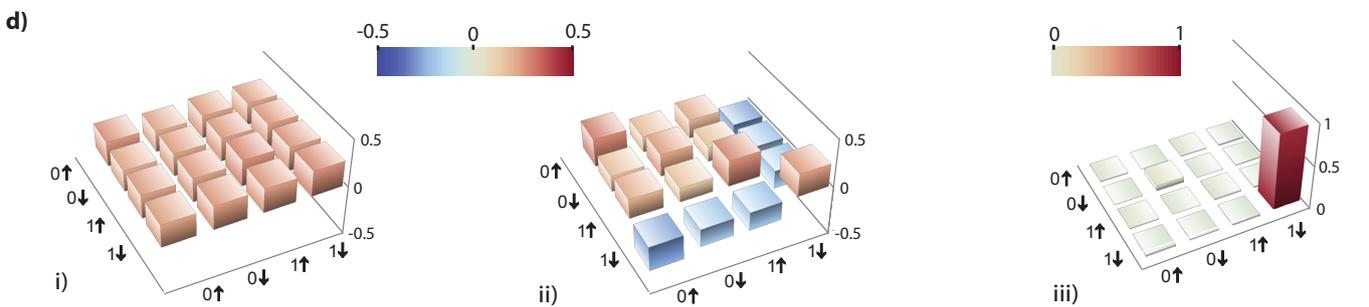

Dobrovitski, V.V., Figure 4